\documentclass[conference]{IEEEtran}
\IEEEoverridecommandlockouts

\usepackage{booktabs}
\usepackage{xcolor,colortbl}
\usepackage{siunitx}
\usepackage{hyperref}
\usepackage{cite}
\usepackage{amsmath,amssymb,amsfonts}
\usepackage{algorithmic}
\usepackage{graphicx}
\usepackage{textcomp}
\usepackage{xcolor}
\usepackage{multirow}

\def\BibTeX{{\rm B\kern-.05em{\sc i\kern-.025em b}\kern-.08em
    T\kern-.1667em\lower.7ex\hbox{E}\kern-.125emX}}
\begin{document}

\title{Device-Robust Acoustic Scene Classification via Impulse Response Augmentation 
\thanks{The LIT AI Lab is supported by the Federal State of Upper Austria. GW's work is supported by the European Research Council (ERC) under the European Union's Horizon 2020 research and innovation programme, grant agreement No 101019375 (Whither Music?).}
}


\author{
\IEEEauthorblockN{Tobias Morocutti$^2$, Florian Schmid$^1$, Khaled Koutini$^2$, Gerhard Widmer$^{1,2}$}
\IEEEauthorblockA{$^1$\textit{Institute of Computational Perception}, $^2$\textit{LIT Artificial Intelligence Lab}  \\
\textit{Johannes Kepler University}, Linz, Austria\\
\{tobias.morocutti, florian.schmid\}@jku.at}
}

\maketitle

\begin{abstract}

The ability to generalize to a wide range of recording devices is a crucial performance factor for audio classification models. The characteristics of different types of microphones introduce distributional shifts in the digitized audio signals due to their varying frequency responses. If this domain shift is not taken into account during training, the model's performance could degrade severely when it is applied to signals recorded by unseen devices. In particular, training a model on audio signals recorded with a small number of different microphones can make generalization to unseen devices difficult. To tackle this problem, we convolve audio signals in the training set with pre-recorded device impulse responses (DIRs) to artificially increase the diversity of recording devices. We systematically study the effect of DIR augmentation on the task of Acoustic Scene Classification using CNNs and Audio Spectrogram Transformers. The results show that DIR augmentation in isolation performs similarly to the state-of-the-art method Freq-MixStyle. However, we also show that DIR augmentation and Freq-MixStyle are complementary, achieving a new state-of-the-art performance on signals recorded by devices unseen during training.

\end{abstract}


\section{Introduction}

Common machine learning theory builds on the assumption that training and test data are drawn from the same underlying distribution~\cite{Vapnik00statistical}. This assumption is often violated in practical applications, causing a performance drop at application time~\cite{Wilson20domainadaptation}. A well-known instance of this problem in the audio domain is the distributional shift caused by the physical characteristics of recording devices~\cite{Heittola20TauDataset}. Digitizing an audio signal using a specific type of microphone results in the device-specific characteristics being encoded into the digitized signal. The problem is particularly severe if the training dataset is recorded using a limited number of recording devices while the application requires the model to generalize to devices unseen during training. 

This problem has been extensively studied in the context of Acoustic Scene Classification (ASC) as part of the annual DCASE challenges~\cite{Heittola20TauDataset, Morato22DcaseTask1}. ASC is the task of predicting an acoustic scene (e.g. traveling by bus, traveling by tram, indoor shopping mall, urban park) given an audio clip. The TAU Urban Acoustic Scenes 2022 Mobile development dataset~\cite{Heittola20TauDataset} used in Task 1 of the DCASE`22 challenge~\cite{Morato22DcaseTask1} uses a Soundman OKM II Classic/Studio A3 as the main recording device referred to as \textit{device A}. The training set further involves two real devices (\textit{device B:} Samsung Galaxy S7, \textit{device C}: GoPro Hero5 Session) and three simulated devices (\textit{S1-S3}). The training subset is imbalanced containing 102,150 samples of device A while other devices are represented by only 7,500 samples. The test subset is balanced across all devices and involves three devices (\textit{S4-S6}) unseen at training time. Previous challenge results~\cite{Heittola20TauDataset, Morato22DcaseTask1} have underlined the difficulty to generalize across different recording devices, showing test accuracy gaps of up to 15 percentage points between device A and unseen devices.

Prior efforts to improve device generalization involve suppressing device-specific information of audio signals~\cite{Koutini20dcasesubmission, Kim22RFN} or increasing the device diversity in the training set by device balancing~\cite{Lee22deviceaware} or augmentation~\cite{Schmid22KD}. As part of the latter, we propose to augment audio signals using a dataset consisting of 66 pre-recorded Impulse Responses (IR) from different microphones. Convolving the audio signals with these device IRs (DIRs) artificially increases the recording device variety in the training dataset.

We validate the performance impact of our proposed DIR augmentation on different architectures including Convolutional Neural Networks (CNNs) and Transformers. With the receptive-field regularized CNN \textit{CP-ResNet}~\cite{Koutini19receptive, Koutini21receptive} and the Patchout faSt Spectrogram Transformer (\textit{PaSST})~\cite{passt}, we choose two state-of-the-art ASC models and evaluate them on the TAU Urban Acoustic Scenes 2020 Mobile development (\textit{TAU20})~\cite{Heittola20TauDataset} and the TAU Urban Acoustic Scenes 2022 Mobile development (\textit{TAU22})~\cite{Heittola20TauDataset} datasets using the official train-test split.

The main contribution of this work is a systematic analysis of the effect of DIR augmentation on the outlined models and datasets. We study the impact of DIR augmentation on the generalization abilities to unseen recording devices and on the overall performance. We compare DIR augmentation to Freq-MixStyle, the device generalization technique that lead to the best accuracies on unseen devices in the ASC task of the DCASE`22 challenge~\cite{Morato22DcaseTask1}. The results show that DIR augmentation consistently improves the generalization performance of models, especially when tested on short clips and in combination with the state-of-the-art method \\ Freq-MixStyle~\cite{Schmid22KD}.\footnote{Code available on GitHub: \href{https://github.com/theMoro/DIRAugmentation}{https://github.com/theMoro/DIRAugmentation}}

\section{Related Work}

\subsection{Recording Device Generalization}
Many different directions and methods have been explored to increase a model's robustness against recording devices. In this regard, Koutini et al.~\cite{Koutini20dcasesubmission} apply the classical Domain Adaptation objectives Maximum Mean Discrepancy~\cite{Gretton12MMD} and Sliced Wasserstein Distance~\cite{Kolouri19SWD} to match recording device distributions and force the model to learn device-invariant high-level embeddings. Primus et al.~\cite{Primus19Exploiting} additionally exploit parallel recordings to create domain-invariant hidden layer representations. Lee et al.~\cite{Lee22deviceaware} balance the device distribution by randomly dropping samples of the dominant \textit{device A}. Device translators are built by Kim et al.~\cite{Kim22translator} to artificially generate samples for underrepresented recording devices. Relaxed instance frequency-normalization is introduced by the same authors~\cite{Kim22RFN} to eliminate device-specific information. Freq-MixStyle~\cite{Kim22RFN, Schmid22KD}, a method that mixes frequency statistics of spectrograms, led to the best generalization performance on unseen devices in the ASC task of the DCASE`22 challenge~\cite{morato21lowcomplexity}.  

\subsection{Impulse Response Augmentation}

Impulse response augmentation is a very common technique to augment audio signals and increase a model's robustness against different acoustic environments by simulating their reverberation. In this regard, Szöke et al.~\cite{Szöke19RIRdataset} publish a dataset of room impulse responses (RIRs) and show that it can be used to improve automatic speech recognition (ASR) when carefully selecting a set of RIRs that match the target environment. Ko et al.~\cite{Ko17RIRspeech} use simulated and real RIRs to create robust models for far-field ASR. Similarly, \cite{Sriram18Robust} and \cite{Ritter16RevRobust} use RIRs to create reverberation robust speech recognition systems.
Compared to RIRs, device impulse responses (DIRs) are less commonly used. Ferr\`as et al.~\cite{Ferras16acousticsimulator} improve speaker-recognition performance by simulating degraded recordings with a variety of augmentation techniques including convolution with DIRs. TADA~\cite{eklund19data}, an audio augmentation pipeline including background noise addition and convolution with RIRs and DIRs, shows to improve a model's robustness against environmental distortions for audio classification tasks.
According to our knowledge, our work is the first detailed study to show that convolution with DIRs drastically improves the generalization to unseen devices for state-of-the-art ASC models.

\section{Device IR Augmentation}

\begin{figure}[h!]
  \centering
  \centerline{\includegraphics[width=\columnwidth]{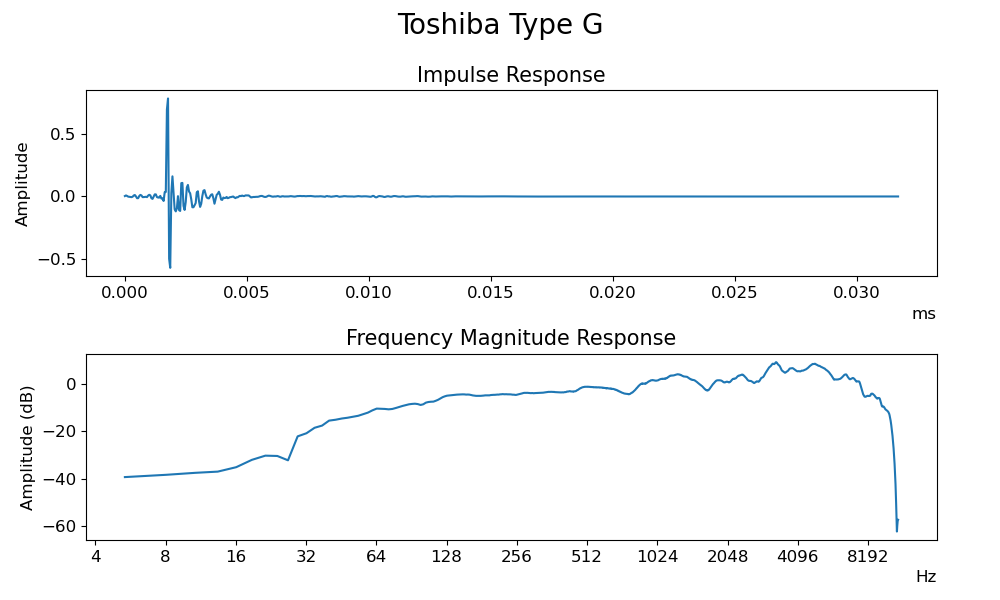}}
  \caption{Impulse response and frequency magnitude response of a \textit{Toshiba Type G} recording device.}
  \label{fig:ir_toshiba}
\end{figure}

As a source of impulse responses, we use 66 freely available DIRs from MicIRP\footnote{\href{http://micirp.blogspot.com}{http://micirp.blogspot.com}}. The DIRs are recorded using a swept-sine method with the microphones placed at a distance of about 20 to 30 cm from the source. MicIRP consists of DIRs of vintage microphones, such as the Toshiba Type G. As the Frequency Magnitude Response in Figure~\ref{fig:ir_toshiba} shows, the devices used have a very specific sonic signature, making them a perfect source for simulating a dataset with high device variability. 

During training, we augment samples from \textit{device A}, the main recording device of the datasets \textit{TAU20}~\cite{Heittola20TauDataset} and \textit{TAU22}~\cite{Heittola20TauDataset}, with a probability of $p_{dir}$ using DIR augmentation. We choose one of the 66 DIRs at random and convolve the waveform with the DIR, cutting off trailing samples to not change the overall length of the waveform. 

\label{sec:illust}

\section{Experimental Setup}

We evaluate our proposed method based on two different datasets and three different models. The two datasets \textit{TAU20}~\cite{Heittola20TauDataset} and \textit{TAU22}~\cite{Heittola20TauDataset} are identical except for the fact that the samples of \textit{TAU20} have a duration of ten seconds, while \textit{TAU22} is split into one-second clips, which results in a more difficult classification task. \textit{TAU20} and \textit{TAU22} are recorded with a limited amount of devices and the official train-test split strongly encourages models to have cross-device generalization capabilities. In the following, we describe the architectures used in the experiments, the audio preprocessing, the training setup and the device generalization techniques we compare our proposed method to.

\subsection{Architectures}

As a first architecture, we use \textit{CP-ResNet}, a receptive-field regularized CNN (RFR-CNN) optimized for ASC~\cite{Koutini19receptive, Koutini21receptive}. CP-ResNet has proven its effectiveness in previous editions of the DCASE ASC challenge~\cite{morato21lowcomplexity, Koutini20dcasesubmission, Koutini21dcasesubmission}. We use CP-ResNet with the number of base channels set to 128, resulting in a model with approximately 4M parameters. Secondly, we build a low-complexity version of the CP-ResNet by reducing the number of blocks and the number of channels per layer. The resulting model has less than 128K parameters and less than 30 million multiply-accumulate operations (MACs), satisfying the complexity requirements for the ASC task of the DCASE`22 challenge~\cite{morato21lowcomplexity}. We denote this model \textit{LC CP-ResNet}. Finally, we use the Audio Spectrogram Transformer \textit{PaSST}~\cite{passt}, which is a complex, self-attention-based model, pre-trained on AudioSet~\cite{audioset2017Gemmeke}, that recently has been shown to achieve state-of-the-art performances on multiple downstream tasks, including ASC~\cite{passt}. PaSST consists of 85M parameters.

\subsection{Audio Preprocessing}

For \textit{CP-ResNet} and \textit{LC CP-ResNet}, we use mono audio with a sampling rate of 22.05kHz. For computing spectrograms, a window size of 2048 samples ($\approx$ 93 ms) and a hop size of 512 samples ($\approx$ 23 ms) is applied. A Mel-scaled filterbank is used to create mel spectrograms with 256 mel bins.
Regarding \textit{PaSST}, we match the AudioSet~\cite{audioset2017Gemmeke} pre-training and use audio at a sampling rate of 32kHz, a hop size of 320 samples (10 ms), a window size of 800 samples (25 ms) and 128 mel bins to compute the mel spectrograms. 

\subsection{Training Setup}

We use AdamW~\cite{loschilov19adamw} as an optimizer with a learning rate schedule including an exponential warm-up, a linear decrease, and a fine-tuning phase to train all models. We use peak learning rates of \num{2e-4} and \num{3e-5} for \textit{CP-ResNet} and \textit{LC CP-ResNet}, respectively, and train both models for 50 epochs using a batch size of 64.
Fine-tuning of \textit{PaSST} on ASC requires only 25 epochs to converge and uses a peak learning rate of \num{1e-5}. We use batch sizes of 80 and 8 for the datasets \textit{TAU22} and \textit{TAU20}, respectively. To avoid overfitting on the downstream task, we apply a structured patchout of 6 on the frequency dimension, similar to~\cite{Schmid22KD}.

\subsection{Device Generalization Methods}

We compare the device generalization abilities of our proposed DIR augmentation method (referenced as \textit{DIR} in short) to Mixup~\cite{Zhang18mixup} and Freq-MixStyle (\textit{FMS})~\cite{Kim22RFN, Schmid22KD}. Mixup creates virtual training samples by linearly mixing two spectrograms and their labels, which has been shown to improve generalization on ASC tasks before~\cite{koutini19mixup}. FMS showed the best generalization capabilities to unseen devices in the latest edition of the ASC task in the DCASE`22 challenge~\cite{Morato22DcaseTask1}. Similar to the original MixStyle~\cite{mixstyle} which mixes channel-wise statistics to improve domain generalization, the goal of FMS is to enhance device generalization by mixing frequency-wise statistics, as the device-style in spectrograms primarily resides in them~\cite{Kim22RFN}. Similar to DIR, FMS is guided by a probability $p_{fms}$ that determines the augmentation strength by specifying the proportion of batches to augment. Additionally, we experiment with using DIR and FMS in combination (\textit{DIR + FMS}). In this case, we first augment the waveforms using the DIRs and apply FMS afterward.

\subsection{Device Generalization Score}
\label{subsec:dg_score}

To quantify device generalization abilities, we introduce the device generalization score (\textit{DG-Score}). We define this measure as the standard deviation of the individual device accuracies. Therefore, a low \textit{DG-Score} indicates a balanced performance of the model across different recording devices.

\begin{table*}[h!]
\begin{tabular}{c|l|ccc|c|ccc|c|ccc|c||c||c}
   & \textbf{Method}  &  \multicolumn{4}{c|}{\textbf{Real Devices}} & \multicolumn{4}{c|}{\textbf{Simulated Devices}} & \multicolumn{4}{c||}{\textbf{Unseen Devices}} & \textbf{Overall} & \textbf{DG}\\ 
   
   \midrule
   
   & & A & B & C & \textbf{Real} & S1 & S2  & S3 & \textbf{Sim} & S4 & S5 & S6 & \textbf{Unseen} & & \\
   \midrule
\parbox[t]{2mm}{\multirow{15}{*}{\rotatebox[origin=c]{90}{TAU22}}} & CP-ResNet &  70.76 & \textbf{63.20} & 63.25 & 65.74 & 56.47 & 52.62 & 57.57 & 55.55 & 51.52 & 48.67 & 44.66 & 48.28 & 56.52 & 7.73 \\

& + Mixup &   70.76 & 62.36 & 62.97 & 65.37 & 58.43 & 54.34 & 59.90 & 57.56 & 53.09 & 50.78 & 48.31 & 50.73 & 57.88 & 6.62 \\

& + DIR &   70.94 & 63.01 & \textbf{64.09} & \textbf{66.02} & 59.42 & 56.54 & 59.29 & 58.42 & 55.70 & 53.46 & 49.47 & 52.88 & 59.10 & 6.00 \\

& + FMS &   \textbf{71.67} & 61.34 & 63.97 & 65.66 & 62.05 & 58.62 & \textbf{63.32} & 61.33 & \textbf{60.02} & 57.85 & 55.91 & 57.93 & 61.64 & 4.34 \\

& \cellcolor{lightgray}+ DIR + FMS &   69.22 & 61.38 & \textbf{64.09} & 64.90 & \textbf{62.49} & \textbf{58.89} & 62.65 & \textbf{61.35} & 59.46 & \textbf{61.20} & \textbf{57.23} & \textbf{59.30} & \textbf{61.84} & \textbf{3.30} \\

\cmidrule{2-16}

& LC CP-ResNet  &   \textbf{67.51} & \textbf{55.15} & 57.36 & 60.01 & 51.73 & 48.30 & 53.19 & 51.07 & 46.86 & 43.25 & 39.28 & 43.13 & 51.40 & 7.86 \\

& + Mixup &   66.21 & 53.92 & 57.17 & 59.11 & 50.71 & 49.39 & 52.93 & 51.01 & 48.02 & 44.70 & 40.27 & 44.33 & 51.48 & 7.04 \\

& + DIR &  66.04 & 53.95 & \textbf{60.65} & \textbf{60.22} & \textbf{54.30} & 49.72 & 54.97 & 53.00 & 48.08 & 49.99 & \textbf{43.44} & 47.17 & 53.46 & 6.40 \\

& + FMS &   66.53 & 54.59 & 57.41 & 59.52 & 50.78 & 49.23 & 54.26 & 51.42 & 48.19 & 47.59 & 41.87 & 45.89 & 52.27 & 6.65 \\

& \cellcolor{lightgray}+ DIR + FMS &   64.98 & 53.76 & 59.95 & 59.57 & 53.55 & \textbf{49.98} & \textbf{55.71} & \textbf{53.08} & \textbf{52.01} & \textbf{52.53} & \textbf{43.44} & \textbf{49.33} & \textbf{53.99} & \textbf{5.71} \\

\cmidrule{2-16}

& PaSST  &   71.75 & \textbf{63.38} & 66.95 & 67.36 & 57.6 & 56.34 & 58.44 & 57.46 & 56.95 & 57.42 & 53.37 & 55.91 & 60.24 & 5.60 \\

& + Mixup &   \textbf{72.00} & 63.36  & 67.69 & \textbf{67.69} & 58.3 & 55.91 & 58.45 & 57.55 & 57.05 & 57.45 & 54.50 & 56.33 & 60.52 & 5.60 \\

& \cellcolor{lightgray}+ DIR &   71.56 & 62.58 & \textbf{67.70} & 67.29 & 60.35 & \textbf{58.62} & \textbf{60.16} & \textbf{59.71} & 59.39 & 59.45 & 56.53 & 58.46 & \textbf{61.82} & 4.55 \\

& + FMS &   71.77 & 62.85 & 66.64 & 67.09 & 60.08 & 57.11 & 59.55 & 58.91 & 58.81 & 58.51 & 56.60 & 57.97 & 61.32 & 4.71 \\

& + DIR + FMS &   70.21 & 62.03 & 66.66 & 66.30 & \textbf{60.96} & 58.30 & 59.78 & 59.68 & \textbf{60.51} & \textbf{59.89} & \textbf{57.25} & \textbf{59.22} & 61.73 & \textbf{3.93} \\

\midrule
\midrule
\parbox[t]{2mm}{\multirow{15}{*}{\rotatebox[origin=c]{90}{TAU20}}} & CP-ResNet & \textbf{81.92} & 69.71 & 75.66 & 75.77 & 68.63 & 65.19 & 70.04 & 67.95 & 66.91 & 60.91 & 56.08 & 61.30 & 68.34 & 7.18 \\

& + Mixup & 80.83 & 72.71 & 75.04 & \textbf{76.19} & 69.41 & 66.89 & 71.70 & 69.33 & 65.56 & 63.90 & 57.66 & 62.37 & 69.29 & 6.42 \\

& + DIR & 79.98 & 72.16 & \textbf{75.85} & 76.00 & 72.46 & \textbf{68.14} & 72.61 & \textbf{71.07} & 68.57 & 67.72 & 60.48 & 65.59 & 70.88 & 5.26 \\

& + FMS & 80.20 & 71.75 & 73.80 & 75.26 & 71.33 & 67.39 & \textbf{73.68} & 70.80 & 69.17 & 68.89 & 65.17 & 67.74 & 71.26 & 4.18 \\

& \cellcolor{lightgray}+ DIR + FMS & 78.18 & \textbf{74.55} & 72.68 & 75.14 & \textbf{72.69} & 66.79 & 71.07 & 70.18 & \textbf{69.37} & \textbf{70.75} & \textbf{66.10} & \textbf{68.74} & \textbf{71.35} & \textbf{3.61} \\
\cmidrule{2-16}

& LC CP-ResNet & 72.30 & 57.00 & 59.98 & 63.10 & 56.73 & 51.92 & 58.42 & 55.69 & 53.27 & 49.90 & \textbf{46.34} & 49.84 & 56.21 & 7.02 \\
& + Mixup & 71.64 & 57.14 & 58.18 & 62.33 & 54.81 & \textbf{55.43} & 60.99 & 57.08 & 55.33 & 52.71 & 43.01 & 50.35 & 56.58 & 7.10 \\
& \cellcolor{lightgray}+ DIR & 70.99 & \textbf{58.76} & \textbf{66.95} & \textbf{65.57} & 59.29 & 52.40 & \textbf{62.81} & \textbf{58.17} & 54.85 & 53.88 & 45.05 & 51.26 & \textbf{58.33} & 7.45 \\
& + FMS & \textbf{73.43} & 56.86 & 60.69 & 63.67 & \textbf{59.49} & 52.99 & 60.40 & 57.63 & 53.92 & 54.97 & 45.66 & 51.52 & 57.60 & 7.12 \\
& + DIR + FMS & 67.11 & 58.38 & 63.81 & 63.10 & 57.86 & 51.43 & 60.12 & 56.47 & \textbf{55.45} & \textbf{58.83} & 44.83 & \textbf{53.04} & 57.53 & \textbf{6.18} \\

\cmidrule{2-16}

& PaSST & 80.12 & \textbf{78.32} & 78.56 & 79.00 & 76.04 & 71.58 & 73.37 & 73.66 & 73.72 & 74.59 & 74.73 & 74.34 & 75.67 & 2.83 \\
& \cellcolor{lightgray}+ Mixup & \textbf{82.30} & 77.95 & \textbf{81.36} & \textbf{80.54} & \textbf{76.57} & 71.68 & 74.34 & \textbf{74.20} & \textbf{75.43} & 74.83 & \textbf{75.03} & \textbf{75.10} & \textbf{76.61} & 3.33 \\
& + DIR & 80.65 & 77.99 & 79.51 & 79.39 & 76.08 & \textbf{71.82} & 74.57 & 74.15 & 74.85 & \textbf{75.45} & 74.75 & 75.02 & 76.18 & 2.71 \\
& + FMS & 79.90 & 77.22 & 77.08 & 78.07 & 74.95 & 71.11 & 72.85 & 72.97 & 74.67 & 73.96 & 74.14 & 74.26 & 75.10 & 2.56 \\
& + DIR + FMS & 78.53 & 76.62 & 78.03 & 77.73 & 75.19 & 71.35 & \textbf{74.91} & 73.82 & 74.91 & 73.98 & 74.08 & 74.32 & 75.29 & \textbf{2.42} \\
\midrule

\end{tabular}
\caption{
Performance on the \textit{TAU Urban Acoustic Scenes 2020 / 2022 Mobile development dataset (\textit{TAU20} / \textit{TAU22})}~\cite{Heittola20TauDataset} following the official test split: Device-wise comparison between baseline, Mixup, Freq-MixStyle (FMS), device impulse response augmentation (DIR) and a combination of DIR and FMS (DIR + FMS). The provided accuracies ($\%$) and device generalization scores (DG) are averaged over three runs and the last five epochs of training. The devices are grouped according to real devices (\textbf{Real}: A, B, C), the seen, simulated devices (\textbf{Sim}: S1, S2, S3) and the unseen, simulated devices (\textbf{Unseen}: S4, S5, S6). For each setup, we mark the best-performing method in gray.
}
\label{tab:device_results}
\end{table*}

\section{Results}

Table \ref{tab:device_results} presents the main results of this paper showing the performance of \textit{CP-ResNet}, \textit{LC CP-ResNet} and \textit{PaSST} on the datasets \textit{TAU20}~\cite{Heittola20TauDataset} and \textit{TAU22}~\cite{Heittola20TauDataset} resulting in six different setups. Each model is evaluated without device generalization methods (\textit{baseline}), with \textit{Mixup}, with \textit{DIR}, with \textit{FMS} and with the combination of DIR and FMS (\textit{DIR + FMS}). Each setup is presented in terms of individual device accuracies, accuracy on all real, simulated and unseen devices, overall accuracy and the DG-Score introduced in Section~\ref{subsec:dg_score}. We refer to the accuracy on the unseen devices as the \textit{unseen accuracy}. The results presented are averages over three independent runs and the last five epochs of training. 

\subsection{Comparison to other Device Generalization Methods}

Table \ref{tab:device_results} shows that experiments including DIR augmentation achieve the highest overall accuracy on five out of the six setups. The results indicate that DIR and FMS are complementary. The combination leads to the highest overall accuracy on three setups and is especially powerful for generalizing to unseen devices, giving the highest unseen accuracy on five out of the six setups and always leading to the lowest DG-Score. Mixup performs best for \textit{PaSST} on \textit{TAU20} both in terms of unseen and overall accuracy, as analyzed in more detail in Section \ref{subsubsec:passt}.

\subsection{Effect on different Models}

\subsubsection{CP-ResNet}

On both datasets, the best-performing single method for CP-ResNet in terms of overall accuracy is FMS. DIR cannot catch up with the performance improvement achieved by FMS but outperforms Mixup and substantially improves upon the baseline increasing the overall accuracy by approximately 2.5 percentage points on both datasets. However, the combination of DIR and FMS improves the overall accuracy slightly compared to using only FMS and increases the unseen accuracy by more than one percentage point on both datasets, leading to the best DG-scores. In comparison to the baseline, DIR + FMS increases the accuracy on the unseen devices by approximately 11 and 7.5 percentage points for the datasets \textit{TAU22} and \textit{TAU20}, respectively. Additionally, using FMS + DIR improves upon the much more complex PaSST model on \textit{TAU22} and outperforms the best PaSST results reported in~\cite{Schmid22KD}, the approach with the best performance in the DCASE`22 challenge.

\subsubsection{Low-Complexity CP-ResNet}

As for the CP-ResNet, the low-complexity network achieves the best performance on both datasets in the setups including DIR. Combining DIR and FMS improves the results on \textit{TAU22} further while DIR alone achieves the best overall accuracy on \textit{TAU20}. DIR + FMS achieves the best results on the unseen devices boosting the accuracy by multiple percentage points on both datasets compared to the baseline. However, we also note that the gain in accuracy on unseen devices comes at the cost of decreasing accuracy on \textit{device A} for some setups and methods and is particularly severe when using the DIR + FMS combination on the LC CP-ResNet. 

\subsubsection{PaSST}
\label{subsubsec:passt}
AudioSet~\cite{audioset2017Gemmeke} is collected from YouTube videos and consists of audio clips recorded by a large variety of different recording devices of diverse quality. Since PaSST is pre-trained on AudioSet, the baseline is already robust against different devices, performing much better than the baselines of the other models that are trained from scratch. In comparison to CP-ResNet and LC CP-ResNet, the performance gap between the baseline and the device generalization methods is much lower for both the overall and unseen accuracies. In the case of \textit{TAU20}, pre-training and fine-tuning clip length (10 seconds) match exactly, allowing PaSST to fully exploit its robustness gained from the pre-training stage. In this scenario, using DIR or FMS can not improve the overall or unseen accuracy. On \textit{TAU22}, consisting of one-second samples, DIR, FMS and the combination of them all enhance the performance with DIR achieving the best overall accuracy.

\subsection{Hyperparameter Sensitivity}

\label{sec:illust}
\begin{figure}[h!]
  \centering
  \centerline{\includegraphics[width=\columnwidth]{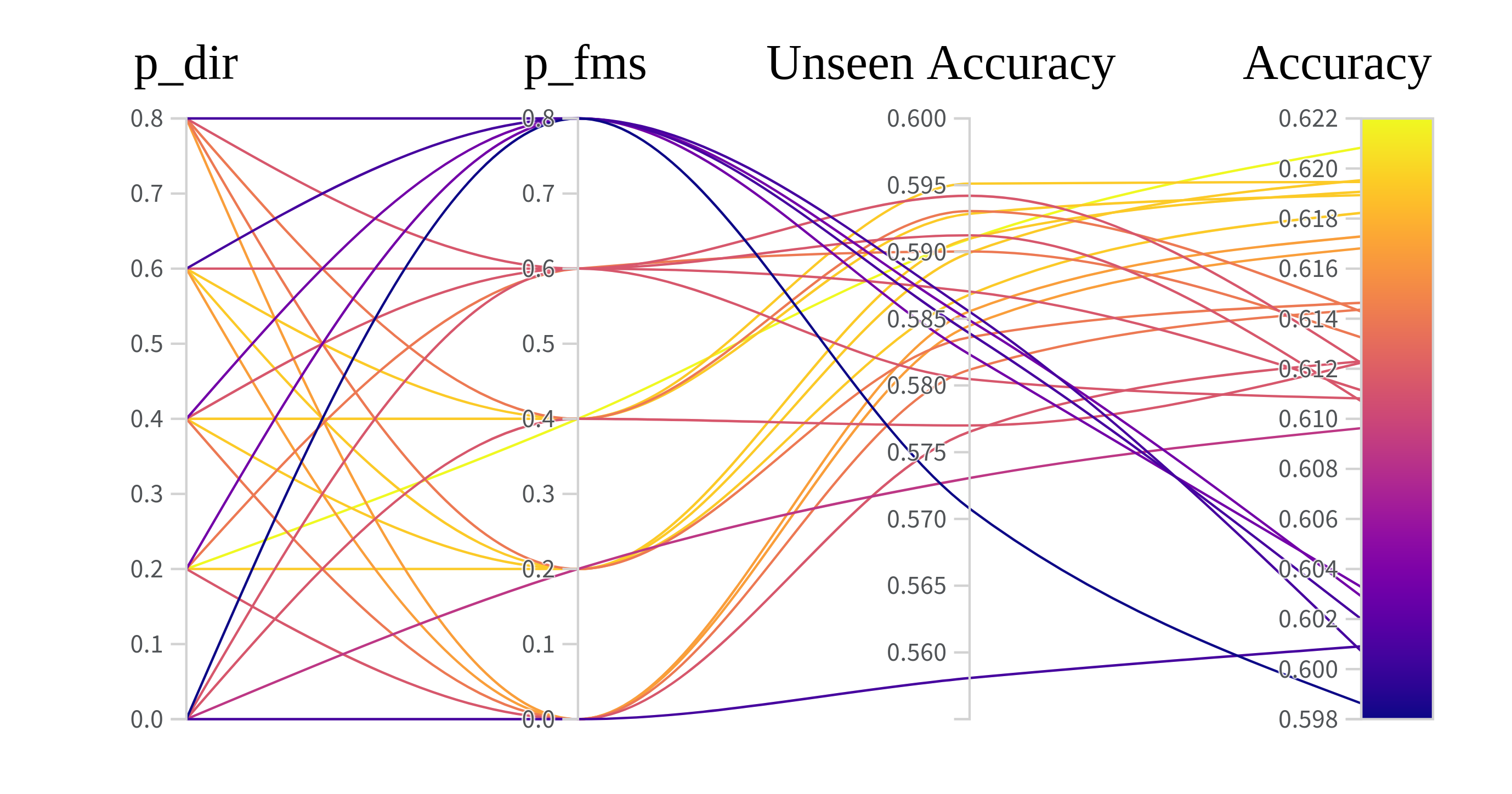}}
  \caption{Parallel coordinate plot visualizing the relationship between the Freq-MixStyle probability ($p_{fms}$), the DIR augmentation probability ($p_{dir}$), accuracy on unseen devices and overall accuracy. Each line is an average over three experiments of running PaSST on TAU22 using DIR + FMS as the device generalization method.}
  \label{fig:parallel_coordinate_plot}
\end{figure}

Relevant hyperparameters for the device generalization methods include the parameters $\alpha$ to determine the shape of the Beta distributions for Mixup and FMS, and $p_{dir}$ and $p_{fms}$ to specify the probabilities to augment a sample and a batch for DIR and FMS, respectively. As suggested in~\cite{Schmid22KD}, $\alpha$ is set to 0.3 and 0.4 for Mixup and Freq-MixStyle, respectively, and $p_{fms}$ takes a value of 0.4. We found that values in the range from 0.4 to 0.6 for $p_{dir}$ yield robust results across models and datasets and decided to use $p_{dir}=0.6$. We choose these values for every setup except for training CP-ResNet on \textit{TAU22} where we experimentally found that values of 0.8 and 0.4 for $p_{fms}$ and $p_{dir}$ give superior results. The hyperparameter configuration is more sensitive when using the combination of DIR and FMS. Figure \ref{fig:parallel_coordinate_plot} visualizes the performance impact of $p_{fms}$ and $p_{dir}$ on the unseen and overall accuracy for PaSST on \textit{TAU22}. The plot reveals that suitable values for $p_{dir}$ are in the range 0.2 to 0.6 and for $p_{fms}$ between 0.2 and 0.4. Choosing either of $p_{dir}$ or $p_{fms}$ too high or setting them to zero decreases performance. This observation holds across all models and datasets.

\section{Conclusion}

In this paper, we address the issue of the distributional shift in digitized acoustic signals caused by different recording device characteristics, which can lead to severe performance drops for audio classification models at application time. We propose device impulse response augmentation as a means to artificially increase the recording device diversity in the training set. Our proposed method drastically improves the device generalization across multiple models and datasets for the task of Acoustic Scene Classification. We compare to other device generalization techniques and conclude that device impulse response augmentation performs similarly to the state-of-the-art method Freq-MixStyle when tested in isolation. However, we also show that device impulse response augmentation and Freq-MixStyle are complementary. The combination is especially powerful to generalize to devices unseen during training, giving the highest accuracy on unseen devices on five out of the six setups tested. 

\bibliographystyle{IEEEtran}
\bibliography{refs}

\begin{thebibliography}{10}
\providecommand{\url}[1]{#1}
\csname url@samestyle\endcsname
\providecommand{\newblock}{\relax}
\providecommand{\bibinfo}[2]{#2}
\providecommand{\BIBentrySTDinterwordspacing}{\spaceskip=0pt\relax}
\providecommand{\BIBentryALTinterwordstretchfactor}{4}
\providecommand{\BIBentryALTinterwordspacing}{\spaceskip=\fontdimen2\font plus
\BIBentryALTinterwordstretchfactor\fontdimen3\font minus
  \fontdimen4\font\relax}
\providecommand{\BIBforeignlanguage}[2]{{%
\expandafter\ifx\csname l@#1\endcsname\relax
\typeout{** WARNING: IEEEtran.bst: No hyphenation pattern has been}%
\typeout{** loaded for the language `#1'. Using the pattern for}%
\typeout{** the default language instead.}%
\else
\language=\csname l@#1\endcsname
\fi
#2}}
\providecommand{\BIBdecl}{\relax}
\BIBdecl

\bibitem{Vapnik00statistical}
V.~Vapnik, \emph{The Nature of Statistical Learning Theory}, ser. Statistics
  for Engineering and Information Science.\hskip 1em plus 0.5em minus
  0.4em\relax Springer, 2000.

\bibitem{Wilson20domainadaptation}
G.~Wilson and D.~J. Cook, ``A survey of unsupervised deep domain adaptation,''
  \emph{{ACM} Trans. Intell. Syst. Technol.}, 2020.

\bibitem{Heittola20TauDataset}
T.~Heittola, A.~Mesaros, and T.~Virtanen, ``Acoustic scene classification in
  {DCASE} 2020 challenge: Generalization across devices and low complexity
  solutions,'' in \emph{Proceedings of the 5th Workshop on Detection and
  Classification of Acoustic Scenes and Events {(DCASE})}, 2020.

\bibitem{Morato22DcaseTask1}
I.~Mart{\'{\i}}n{-}Morat{\'{o}}, F.~Paissan, A.~Ancilotto, T.~Heittola,
  A.~Mesaros, E.~Farella, A.~Brutti, and T.~Virtanen, ``Low-complexity acoustic
  scene classification in {DCASE} 2022 challenge,'' in \emph{Proceedings of the
  7th Workshop on Detection and Classification of Acoustic Scenes and Events
  ({DCASE})}, 2022.

\bibitem{Koutini20dcasesubmission}
K.~Koutini, F.~Henkel, H.~Eghbal-zadeh, and G.~Widmer, ``{{CP-JKU} Submissions
  to {{DCASE}}’20: Low-Complexity Cross-Device Acoustic Scene Classification
  with RF-Regularized {CNNs}},'' \emph{Tech. Rep., DCASE Challenge}, 2020.

\bibitem{Kim22RFN}
B.~Kim, S.~Yang, J.~Kim, H.~Park, J.~Lee, and S.~Chang, ``Domain generalization
  with relaxed instance frequency-wise normalization for multi-device acoustic
  scene classification,'' in \emph{23rd Annual Conference of the International
  Speech Communication Association {(Interspeech)}}.\hskip 1em plus 0.5em minus
  0.4em\relax {ISCA}, 2022.

\bibitem{Lee22deviceaware}
J.-H. Lee, J.-H. Choi, P.~M. Byun, and J.-H. Chang, ``Multi-scale architecture
  and device-aware data-random-drop based fine-tuning method for acoustic scene
  classification,'' in \emph{Proceedings of the 7th Workshop on Detection and
  Classification of Acoustic Scenes and Events ({DCASE})}, 2022.

\bibitem{Schmid22KD}
F.~Schmid, S.~Masoudian, K.~Koutini, and G.~Widmer, ``Knowledge distillation
  from transformers for low-complexity acoustic scene classification,'' in
  \emph{Proceedings of the 7th Workshop on Detection and Classification of
  Acoustic Scenes and Events {(DCASE)}}, 2022.

\bibitem{Koutini19receptive}
K.~Koutini, H.~Eghbal{-}zadeh, M.~Dorfer, and G.~Widmer, ``The receptive field
  as a regularizer in deep convolutional neural networks for acoustic scene
  classification,'' in \emph{27th European Signal Processing Conference
  {(EUSIPCO)}}.\hskip 1em plus 0.5em minus 0.4em\relax {IEEE}, 2019.

\bibitem{Koutini21receptive}
K.~Koutini, H.~Eghbal{-}zadeh, and G.~Widmer, ``Receptive field regularization
  techniques for audio classification and tagging with deep convolutional
  neural networks,'' \emph{{IEEE} {ACM} Trans. Audio Speech Lang. Process.},
  2021.

\bibitem{passt}
K.~Koutini, J.~Schl{\"{u}}ter, H.~Eghbal{-}zadeh, and G.~Widmer, ``Efficient
  training of audio transformers with patchout,'' in \emph{23rd Annual
  Conference of the International Speech Communication Association
  {(Interspeech)}}.\hskip 1em plus 0.5em minus 0.4em\relax {ISCA}, 2022.

\bibitem{Gretton12MMD}
A.~Gretton, K.~M. Borgwardt, M.~J. Rasch, B.~Sch{\"{o}}lkopf, and A.~J. Smola,
  ``A kernel two-sample test,'' \emph{J. Mach. Learn. Res.}, 2012.

\bibitem{Kolouri19SWD}
S.~Kolouri, K.~Nadjahi, U.~Simsekli, R.~Badeau, and G.~K. Rohde, ``Generalized
  sliced wasserstein distances,'' in \emph{Advances in Neural Information
  Processing Systems 32 {(NeurIPS)}}, 2019.

\bibitem{Primus19Exploiting}
P.~Primus, H.~Eghbal{-}zadeh, D.~Eitelsebner, K.~Koutini, A.~Arzt, and
  G.~Widmer, ``Exploiting parallel audio recordings to enforce device
  invariance in cnn-based acoustic scene classification,'' in \emph{Proceedings
  of the 4th Workshop on Detection and Classification of Acoustic Scenes and
  Events {(DCASE})}, 2019.

\bibitem{Kim22translator}
B.~Kim, S.~Yang, J.~Kim, and S.~Chang, ``{QTI} submission to {DCASE} 2021:
  residual normalization for device-imbalanced acoustic scene classification
  with efficient design,'' \emph{Tech. Rep., DCASE Challenge}, 2021.

\bibitem{morato21lowcomplexity}
I.~Mart{\'{\i}}n{-}Morat{\'{o}}, T.~Heittola, A.~Mesaros, and T.~Virtanen,
  ``Low-complexity acoustic scene classification for multi-device audio:
  Analysis of {DCASE} 2021 challenge systems,'' in \emph{Proceedings of the 6th
  Workshop on Detection and Classification of Acoustic Scenes and Events
  ({DCASE})}, 2021.

\bibitem{Szöke19RIRdataset}
I.~Sz{\"{o}}ke, M.~Sk{\'{a}}cel, L.~Mosner, J.~Paliesek, and J.~H.
  Cernock{\'{y}}, ``Building and evaluation of a real room impulse response
  dataset,'' \emph{{IEEE} J. Sel. Top. Signal Process.}, 2019.

\bibitem{Ko17RIRspeech}
T.~Ko, V.~Peddinti, D.~Povey, M.~L. Seltzer, and S.~Khudanpur, ``A study on
  data augmentation of reverberant speech for robust speech recognition,'' in
  \emph{{IEEE} International Conference on Acoustics, Speech and Signal
  Processing {(ICASSP)}}.\hskip 1em plus 0.5em minus 0.4em\relax {IEEE}, 2017.

\bibitem{Sriram18Robust}
A.~Sriram, H.~Jun, Y.~Gaur, and S.~Satheesh, ``Robust speech recognition using
  generative adversarial networks,'' in \emph{{IEEE} International Conference
  on Acoustics, Speech and Signal Processing {(ICASSP)}}.\hskip 1em plus 0.5em
  minus 0.4em\relax {IEEE}, 2018.

\bibitem{Ritter16RevRobust}
M.~Ritter, M.~M{\"{u}}ller, S.~St{\"{u}}ker, F.~Metze, and A.~Waibel,
  ``Training deep neural networks for reverberation robust speech
  recognition,'' in \emph{Proceedings of the 12th {ITG} Symposium on Speech
  Communication}.\hskip 1em plus 0.5em minus 0.4em\relax {IEEE}, 2016.

\bibitem{Ferras16acousticsimulator}
M.~Ferras, S.~R. Madikeri, P.~Motl{\'{\i}}cek, S.~Dey, and H.~Bourlard, ``A
  large-scale open-source acoustic simulator for speaker recognition,''
  \emph{{IEEE} Signal Process. Lett.}, 2016.

\bibitem{eklund19data}
V.-V. Eklund, ``Data augmentation techniques for robust audio analysis,''
  \emph{Master's Thesis, Tampere University}, 2019.

\bibitem{Koutini21dcasesubmission}
K.~Koutini, S.~Jan, and G.~Widmer, ``{CPJKU Submission to DCASE21: Cross-Device
  Audio Scene Classification with Wide Sparse Frequency-Damped {CNNs}},''
  \emph{Tech. Rep., DCASE Challenge}, 2021.

\bibitem{audioset2017Gemmeke}
J.~F. Gemmeke, D.~P.~W. Ellis, D.~Freedman, A.~Jansen, W.~Lawrence, R.~C.
  Moore, M.~Plakal, and M.~Ritter, ``Audio set: An ontology and human-labeled
  dataset for audio events,'' in \emph{{IEEE} International Conference on
  Acoustics, Speech and Signal Processing {(ICASSP)}}.\hskip 1em plus 0.5em
  minus 0.4em\relax {IEEE}, 2017.

\bibitem{loschilov19adamw}
I.~Loshchilov and F.~Hutter, ``Decoupled weight decay regularization,'' in
  \emph{7th International Conference on Learning Representations
  {(ICLR)}}.\hskip 1em plus 0.5em minus 0.4em\relax OpenReview.net, 2019.

\bibitem{Zhang18mixup}
H.~Zhang, M.~Ciss{\'{e}}, Y.~N. Dauphin, and D.~Lopez{-}Paz, ``mixup: Beyond
  empirical risk minimization,'' in \emph{6th International Conference on
  Learning Representations {(ICLR)}}.\hskip 1em plus 0.5em minus 0.4em\relax
  OpenReview.net, 2018.

\bibitem{koutini19mixup}
K.~Koutini, H.~Eghbal-zadeh, and G.~Widmer, ``{CP-JKU} submissions to
  {{DCASE}}’19: Acoustic scene classification and audio tagging with
  receptive-field-regularized {CNN}s,'' \emph{Tech. Rep., DCASE Challenge},
  2019.

\bibitem{mixstyle}
J.~Fu, Y.~Zhong, and F.~Yang, ``Adversarial domain generalization with
  mixstyle,'' in \emph{International Conference on Advanced Robotics and
  Mechatronics {(ICARM)}}.\hskip 1em plus 0.5em minus 0.4em\relax {IEEE}, 2022.

\end{thebibliography}

\end{document}